GLEX-21-5.1.6.x62197

# An L-class Multirole Observatory and Science Platform for Neptune

**James McKevitt[a*], Sophie Bulla[b], Tom Dixon[c], Franco Criscola[d], Jonathan Parkinson-Swift[e], Christina Bornberg[f], Jaspreet Singh[g], Kuren Patel[h], Aryan Laad[i], Ethan Forder[c], Louis Ayin-Walsh[j], Shayne Beegadhur[c], Paul Wedde[b], Bharath Simha Reddy Pappula[k], Thomas McDougall[c], Madalin Foghis[c], Jack Kent[c], James Morgan[j], Utkarsh Raj[g], Carina Heinreichsberger[a]**

[a] *Institute of Astrophysics, University of Vienna, Türkenschanzstrasse 17, 1180 Wien, Austria*
[b] *Conceptual Exploration Research, Germany*
[c] *Loughborough University, Epinal Way, Loughborough LE11 3TU, United Kingdom*
[d] *Embry-Riddle Aeronautical University, 1 Aerospace Boulevard, FL 32114-3900, United States*
[e] *Nottingham Trent University School of Science and Technology, NG11 8NS, Nottingham, United Kingdom*
[f] *University of Applied Science Technikum Wien, Höchstädtplatz 6, 1200 Wien, Austria*
[g] *Conceptual Exploration Research, India*
[h] *Conceptual Exploration Research, United States*
[i] *University of the Arts London, Central Saint Martins, 1 Granary Square, London N1C 4AA, United Kingdom*
[j] *Conceptual Exploration Research, United Kingdom*
[k] *Moscow Aviation Institute, Moscow, 125080, Russian Federation*
* Corresponding Author

**Abstract**

A coming resurgence of super heavy-lift launch vehicles has precipitated an immense interest in the future of crewed spaceflight and even future colonisation efforts. While it is true that a bright future awaits this sector, driven by commercial ventures and the reignited interest of old space-faring nations, and the joining of new ones, little of this attention has been reserved for the science-centric applications of these launchers. The Arcanum mission is a proposal to use these vehicles to deliver an L-class observatory into a highly eccentric orbit around Neptune, with a wide-ranging suite of science goals and instrumentation tackling Solar System science, planetary science, Kuiper Belt Objects and exoplanet systems.

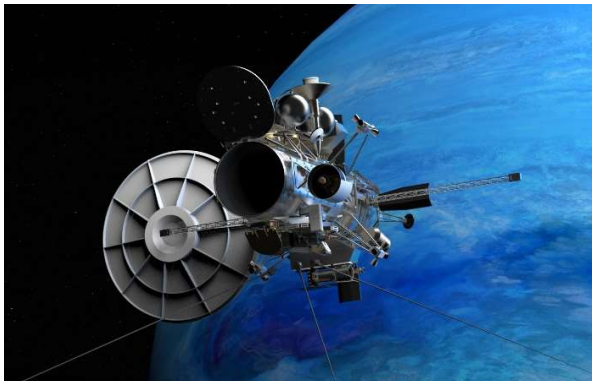

Fig. 1. The Arcanum Mission in orbit around Neptune

## 1. Introduction

We are undeniably entering a new age of space exploration, where access to the sector by both governments and private companies is unprecedented. Tremendous private funding and a resurgence in the appetite of governments to invest in prestige enhancing programmes means there is a demand for launchers and support for their development. With access to space heavily influenced by national security concerns and funding, and with a returning focus on this domain in defence circles, coupled with a rapidly growing space economy and paths to profitability in the sector, a plethora of routes to space are being presented, allowing increased redundancy and annual lifting capacity.

One result of this new 'race to space' is the development of a new generation of super heavy-lift launch vehicle (SHLLV), these being rockets capable of delivering over 50 tonnes to low Earth orbit (LEO). These vehicles are nothing new, with successful examples flown over 50 years ago in the form of the Saturn V and Energia. The former was designed specifically and for the sole purpose of crewed Lunar exploration, and the latter served to support the Buran spacecraft to orbit and launch the Polyus spacecraft.

As we see this resurgence of such large rockets, a lively discussion is taking place around their use within the context of crewed spaceflight, with significant levels of both private and government funding supporting these efforts. Presented in this paper, however, is the case for a large single science-centric payload to take advantage of this capacity.

Arcanum is a large strategic science mission – sometimes known as an L-class or Flagship mission –

consisting of a multirole orbiter-lander spacecraft, with primary science goals surrounding Neptune, Triton, and Kuiper Belt observations. When at its destination in a highly-eccentric Neptune orbit, the orbiter will use a 1-metre diameter telescope to observe Kuiper Belt Objects and exoplanets. Simultaneously, other instrumentation on the orbiter will answer questions about Neptune's magnetosphere and atmosphere, working in tandem with two landing penetrators and a soft-landing probe, which will descend to Triton, answering questions about its composition, structure and evolution.

## 2. Context
### 2.1 Industry

As aforementioned, one of the driving factors behind the current revolution in available launchers is the evolution of private industry. With aerospace companies traditionally taking the role of contractor, such as for the previous generation of SHLLV, a change in appetite for private investment in the sector and support of this new initiative through government funding has meant the growth of a large number of private companies building and operating launch vehicles. A natural progression for these companies has now seen them offering not only the service of launch but also services through hardware they also operate in space.

The foundations for this so-called 'billionaire space race' were arguably laid as early as the 1990s by visionary Peter Diamandis creator of, amongst numerous other revolutionary ventures, the X PRIZE Foundation. This made its first award, the Ansari X PRIZE, in 2004 to the predecessor of SpaceShipTwo, now finally within months of its targeted commercial flights with Virgin Galactic. Virgin Galactic is only one of the numerous commercial ventures promising to make space more accessible, currently most comparable to Blue Origin in this area, who are targeting 20th July for their maiden passenger flight. Blue Origin is further involved in the heavier-lift launcher market, relevant to this study and further discussed later, through their staged-combustion BE-4 engine, planned for flight on the United Launch Alliance's Atlas V replacement, the Vulcan Centaur. These companies are intrinsically driven by competition, growth and profit, some of the reasons for their success relative to government programmes. However, companies such as SpaceX, again relevant to the Arcanum mission and so discussed later, and Rocket Lab are both operated by visionary leaders, still driven by profit, but with the ambition to bring about large changes in humanity's understanding of space. Rocket Lab's Peter Beck, for example, has indicated his strong interest in operating dedicated deep space scientific spacecraft, an exciting revolution for the space science community attempting in any way to connect themselves with this heavily industrialised revolution. This also presents a welcome change of attitude for the scientific community from large, privately-funded industry ventures, which have been shown to damage scientific observations and the safety of Earth orbit.

### 2.2 Science

Missions to the outer planets are regularly investigated, and numerous working groups in the planetary science community have previously advocated in detail for these with complete science cases and spacecraft proposals. Neptune features heavily amongst these, and innovative mission proposals present comprehensive and compelling cases for their funding [1,2]. Large scale roadmaps created by national space agencies, intended to provide direction for the entire space science community also provide detailed lists of objectives at these targets [3]. Interest is high given the lengthy period since a previous mission, with the Neptunian System receiving only one in-situ visit by Voyager. The closest of these missions to approval, although very recently passed over by NASA [4] is the Trident spacecraft [5], focusing even more specifically on Triton, Neptune's major moon.

### 2.3 Launchers

The Arcanum Mission is not only an example of what large strategic science missions can deliver to the scientific community, as it also aims to set a new precedence in the use cases of SHLLVs. At its current design iteration, the spacecraft and interplanetary propulsion require a launch vehicle capable of accommodating a 7-meter diameter payload, with a height of 17 meters and a mass of over 45 tonnes. Three main launchers were considered for transporting Arcanum:

- NASA's Space Launch System (SLS)
- SpaceX's Starship
- Blue Origin's New Glenn.

These launch vehicles were predominantly selected based on their payload volume and mass carrying characteristics, along with their deemed high likelihood of flight.

#### 2.3.1 Space Launch System

SLS has the capability to launch crewed and un-crewed missions, and usefully for Arcanum, offers additional adaptability in its upper stage with configurations for a range of requirements [6]. The SLS Block 2 Cargo variant is the closest launcher capable of transporting Arcanum on its deep-space transfer, chosen over the SLS Block 1B Cargo variant due to inadequate height of payload fairing and mass carrying capability of 38 tonnes (83,700 lbs.). Block 2 cargo also offers limits margins with a capacity of 46 tonnes (101,400 lbs.). The quoted payload masses and construction of these additional variants are highly dependent on the performance of SLS Block 1, slated for launch in late 2021. Integrating with SLS also offers the possibility for Arcanum to exchange its currently envisaged boost stages, with the Exploration Upper Stage (EUS), powered by four RL10 engines. This would offer capabilities currently integrated into the Arcanum spacecraft, such as orbital manoeuvring and station keeping. However, this very much remains tied to launcher selection.

SLS still poses some uncertainty due to its high cost per launch, estimated to be $2 billion per flight [7], perhaps a consequence of its lack of full or partial reusability. This poses some concerns as to its viability as a long term competitive launch vehicle that can rival those of the private sector. As Arcanum is not intended to be launched for at least 15 years, this long term uncertainty is critical.

*2.3.2 New Glenn*

New Glenn is Blue Origin's next-generation rocket with super heavy-lift capabilities, with a focus on 'fault tolerance, safety, and reusability' [8]. New Glenn offers partial reusability of its system in that after 1$^{st}$ stage separation, the stage performs a reorientation manoeuvre followed by a landing burn and touchdown on a landing platform at sea. This partial reusability can be expected to make New Glenn more economically competitive than SLS, however, the clandestine business model of Blue Origin makes this almost impossible to estimate.

Powered by Blue Origin's own BE-4 engines, New Glenn is expected to deliver 45 tonnes to LEO [8]. For New Glenn to be considered as a launch provider for Arcanum, tight margins would be placed on the vehicle and its interplanetary propulsion system.

*2.3.3 Starship*

SpaceX's Starship launcher has been designed from the ground up to be a fully reusable transport system that can be utilised to deliver payloads to LEO, the Moon, Mars and deep space [9]. The vehicle is currently under development, as are SLS and New Glenn, although under a more public iterative design process, aiming to 'evolve rapidly to meet near term and future customer needs' [9]. As with SLS, Starship is offered in multiple configurations: Crewed Deep Space, Crewed Earth Point to Point, Cargo, Lunar HLS and Tanker. It is likely the simplest of these will be offered first through Starship Cargo and Tanker missions.

SpaceX has been the recipient of NASA funding for a number of programmes, including their successful operation of crewed missions to the International Space Station and more recently a full-scale orbital propellant transfer demonstration was recently awarded $53 million [10]. It is expected that if this capability is included on Starship and that a fully expendable configuration is used that tremendous payloads of 100 tonnes could be delivered to deep space.

This large range of supported payload masses are beneficial to a vehicle currently under design, but with the need for a payload volume constraint, such as Arcanum. This available volume on Starship is also relatively large, offering an 8-meter diameter fairing and an optional extended volume for payloads of up to 22 m in height [9]. This results in the largest usable payload volume of any current or development launcher.

Regarding financial competitiveness, SpaceX claims that Starship could cost as little as $2 million per flight, a considerable reduction from that of SLS. While such high specifications for such a decreased cost seems questionable, the growing track record of safe and reliable space operations offered by SpaceX serves to answer any concerns. Furthermore, growing confidence in Starship's viability can be most recently seen in NASA's awarding of a $2.9 billion Human Landing System (HLS) contract to configure Starship for Artemis Program lunar landings [11]. The U.S. Air Force has also proposed investing $50 million in Starship for Earth point-to-point cargo delivery [12].

The growing confidence that SpaceX can provide an operational Starship for such low cost, along with the view that its operation is somewhat future-proofed has led to its selection as the launch vehicle for the Arcanum mission.

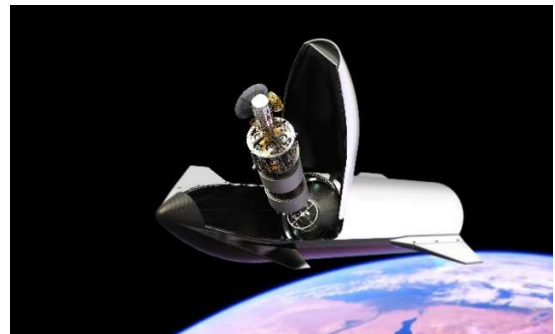

Fig. 2. Arcanum deployed by Starship

**3. Science Objectives**

The Arcanum Mission is designed around science goals for a diverse range of targets, wholly fulfilling its 'multirole' designation. Such a wide-ranging instrument suite will, of course, always be suitable for a large number of scientific objectives, and with instrument data released in the standard open manner, these are not limited to those stated here, thanks to the ingenuity of space science researchers. Those listed are intended to show the range of tasks for which the Arcanum mission is applicable, provide a justification for the selected instruments, and are clearly defined around:

- Neptunian System
- Kuiper Belt Objects
- Solar System
- Exoplanets

The mission begins at launch and continues throughout the transfer, Neptune orbit, and ends with mission disposal. This reflects the gathering of data during all phases of the mission aimed at addressing the science goals defined here.

Following separation from propulsion stages, the remaining spacecraft is placed in a highly eccentric Poseidocentric orbit, this being an orbit around Neptune. This will be at an inclination of around 50 degrees, allowing some polar coverage of Neptune with potential for increase, while not complicating access to Neptune's moons, in particular Triton.

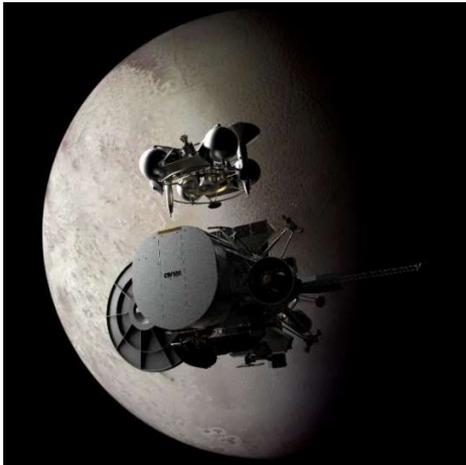

Fig. 3. Somerville releases Bingham at Triton

This spacecraft consists of two key components
- Somerville: Neptune Orbiter
- Bingham: Triton Lander

further divided with the Bingham lander, destined for Triton, consisting of
- Central soft landing probe
- Two surface penetrators

The two key components, Somerville and Bingham, are named after Mary Somerville, an eighteenth and nineteenth-century Scottish polymath and first female member of the Royal Astronomical Society alongside Caroline Herschel, and Hiram Bingham III, nineteenth and twentieth-century American explorer.

Most writing here is concerned with Somerville, and therefore atmospheric planetary science, Solar System and exoplanet science. Additional details on Bingham and the surface penetrators will be published in subsequent conference papers.

*3.1 Neptune*

Planetary atmospheres are dynamic systems, interacting with the solid body via processes such as subduction and outgassing, and escaping through atmospheric erosion by stellar winds or photochemistry [13]. Discussed here under the term 'atmospheric loss', also known as 'atmospheric escape', this phenomenon is key to understanding the dynamics of any system with an atmosphere. This concept is also mentioned in the context of Triton and exoplanet science in later sections, with mechanisms identical but measurement techniques varied. A successful and dedicated mission to the observation of this atmospheric loss is MAVEN (Mars Atmosphere and Volatile Evolution Mission), a NASA mission investigating Mars' atmosphere through a payload of eight science instruments [14]. When solar wind hits the atmosphere of Mars, a shockwave forms as dense solar wind plasma where the number of charged particles is enhanced. The solar wind then interacts with the atmosphere and ions are created. At the same time, an electric field is generated pointing away from the planet, causing ions to move along electric field lines and be lost. Approximately 25% of the ions are lost this way, with the other 75% drifting around Mars and escaping in the solar wind direction. A general loss rate can be estimated by mapping the regions of high and low ion escape. However, this only provides a lower limit as neutrals are not included in these measurements.

Within the Particles and Fields package on MAVEN several instruments specialize in solar wind measurements, two of these being a Solar Wind Electron Analyzer (SWEA) [15] and Solar Wind Ion Analyzer (SWIA) [16]. These instruments serve as a basis for a similar instrument package on Somerville. SWEA produces an energy spectrum, selecting electrons with energies within the region where most atoms ionize in planetary atmospheres. These spectra provide information about the distribution of energy fluxes, giving insight into the loss and ionization of species in the atmosphere. SWIA compliments SWEA through measurements of the solar wind around Mars as well as magnetosheath proton flow.

Additional relevant instruments such as the NGIMS (Neutral gas and Ion spectrometer) help complete the picture by working closely together with IUVIS (Imaging Ultraviolet Spectrometer) and LPW (Langmuir Probe and Waves) to measure the composition and isotopic ratios of the upper atmosphere [17].

The actual escape rates of the species however can only be calculated with models, using the measurements provided by these instruments, such as done by Lillis [18] using MAVEN data.

*3.1.2 Triton*

Triton is one potential source of interest when discussing habitability in the Solar System, and as such has benefitted from a number of innovative concepts aiming to better understand its environment. Somerville's atmospheric science suite will work in a similar way during fly-bys high over Triton, hoping to understand the height to which this weak atmosphere extends.

Regarding Triton surface science, a detailed description of this is reserved for future papers more dedicated to Bingham. However, one area presenting a new approach only recently more adapted in space is Raman spectroscopy. The analysis of molecular spectroscopy is a vital part of developing our understanding during the exploration of the neighbouring planets of the Solar System. This is especially important when investigating factors such as mineralogy, atmospheric aerosols and the presence of life. Infrared and Raman Spectrometers are becoming more prevalent in payloads that aim to analyse molecular samples on account of their energy sensitive methods which can reveal the specific 'signature' of present compounds, alongside their components and properties. This can be seen in some of the more recent ongoing and future mission proposals due to the development of the miniaturisation and optimisation of the instruments in recent years [19,20].

One of the advantages of applying Raman is that it is a rapid and non-intrusive scattering technique that is already used to examine the structural identity of rock and mineral samples on Earth. It is desired alongside IR due to its ability to analyse samples involving water, which often causes interference issues during IR analysis. Raman also has the proven advantage of being able to distinguish water in its different forms, something beneficial when studying Triton's icy regolith [21].

The use of laser-induced Raman spectrometry in missions such as Mars2020 (SHERLOC [22] and SuperCam [23]) and ExoMars (RLS) [24] attests to this whilst using several different Raman techniques including Deep UV resonance and time-resolved resonance Raman. The success of these instruments will lead the discussion of adapting the CIRS instrument for the Europa Lander mission concept [25], especially as ExoMars will be the pioneer for using such instruments.

These examples, accompanied by demonstrated research on samples obtained from the Murchison, Allende and Itokawa asteroids [26], signify the practicability of the technique and the necessity of instruments that can operate in Raman specific modes for future studies to characterise atmospheric, surface and subsurface materials present on astronomical bodies [21,27].

Considering the above, Raman Spectroscopy compliments the mission goal appropriately and will have a large impact on the future of planetary exploration in general. Raman instruments included on the landing probe's payload will allow us to study the composition of both the surface and subsurface materials without damaging the area of interest. Equipment adapted from CIRS, which is currently being adapted for the similar icy environment of Europa, would be ideal for this purpose.

Furthermore, there is also the potential to analyse atmospheric aerosols based on the location of the lander, where objects of interest such as the volcanic plumes are able to assist in the understanding of Triton's thin atmosphere, volcanic activity and seasonal heating.

*3.1.2 Kuiper Belt Objects*

The clustering of KBOs, indicative of a large mass in the outer Solar System [28], was one of the initially enthusing elements at the inception of this mission. This still provides a solid science goal and use for the large telescope component of Somerville, and should therefore receive science time. In addition, the study of KBOs themselves is a neglected field and further dedicated time of their study, using a telescope unobscured by zodiacal light, discussed in the proceeding section, would be beneficial to the field.

*3.1.3 Solar System*

Zodiacal light (ZL) is caused by solar radiation scattered by interplanetary dust particles (IDP) and contaminates observations in all wavelengths. The optical and infrared bands are most affected [29–32] and a correct model is, therefore, essential for all research addressing questions about the outer Solar System and beyond [33].

However, modelling ZL is very challenging, the main uncertainty being the photometrical properties of IDP such as thermal emissivity or the effect of background stars on the optical wavelength regime [31].

Shannon et al. [34] show the effect of ZL on four example observations, illustrating that the signal-to-noise ratio is strongly impacted if the zodiacal light has a similar surface brightness as the observed object. This limits detection possibilities based on surface brightness.

The placement of a spacecraft outside the ZL region would be beneficial not only for the constraint of ZL properties, and therefore IDP, but also for 'clean', less obscured observations of objects beyond the Solar System.

*3.1.4 Exoplanets*

The understanding of exoplanet atmospheres is of great interest in and of itself. When considered through the lens of exoplanet habitability, however, their properties and more specifically their composition becomes of high importance. It is widely accepted that with current and near-future technology, life could at the very least only be detected on an exoplanet if it possesses an atmosphere. Atmospheric escape, aforementioned in the context of Neptune in Section 3.1.1, is, therefore, one of the key questions for habitability. To study this effect, different techniques can be applied given the limits of remote sensing. One promising, and well-practised technique currently performed by the Hubble Space Telescope (HST) and the Spitzer Space Telescope is transit spectroscopy. Here, the host star emits photons of varied wavelengths according to spectral type. Some of these photons are then absorbed by the atmosphere of the orbiting exoplanet, depending on the photon wavelength and atmospheric composition. The resulting transmission spectrums can be used to determine atmospheric compositions and even constrain atmospheric loss. When a planet suffers atmospheric loss, gases are distributed around the planet creating an envelope that is orders of magnitudes larger than its radius. Lighter species are lost more easily, such as hydrogen. As hydrogen absorbs at a different wavelength, several lines can be used as tracers. Presently, the Lyman-alpha line at 121.567nm can be used for exoplanets at distances up to 19pc [35,36]. Signals that come from further away cannot be detected, as the interstellar medium (ISM), consisting of hydrogen, absorbs this line [37]. An additional parameter that is needed to study atmospheric loss is the Roche lobe, the region around an object in which material is gravitationally bound and outside of which everything is lost to space. If the envelope traced by Lyman-alpha is larger than this region material is lost to space [36]. By measuring the blue and redshift of the line, the velocity of the lost particles can be estimated and therefore a loss rate can be determined.

**4. Mission Design**

Arcanum Mission architecture and key mission elements can be seen in Fig. 4, and have been defined according to standard practice [38]. These helped guide the mission study, with somewhat of an asymmetrical approach taken given the setting of the launch segment and mission subject. Furthermore, the initial study's scope has dictated attention on Somerville. Further details on the operation of Bingham and the penetrators will be made available in subsequent conference papers.

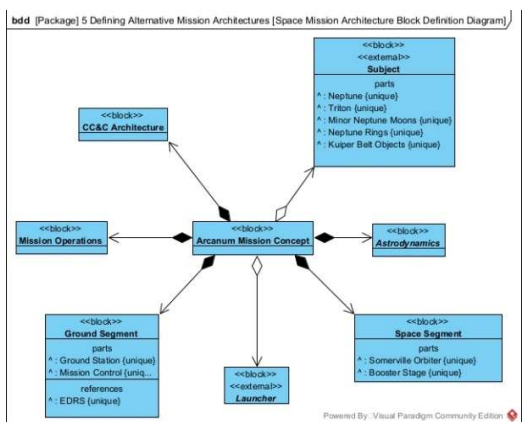

Fig. 4. Arcanum Space Mission Architecture

*4.1 Transfer*

Neptune orbits at a distance of 30 AU from the Sun, meaning a short-duration direct transfer would be a very fuel-inefficient trajectory. Fig. 5 shows a direct transfer from Earth to Neptune. Using a trajectory optimization algorithm, it can be found that an approximate 180 km$^2$/s$^2$ of launch energy (C3) must be added to the spacecraft in its parking orbit to achieve a direct transfer within 10-15 years, an amount no current launch vehicle can provide. That means a 9-10 km/s departure ΔV increase from a 300km parking orbit. The required energy will change depending on the desired transfer time, with energy increasing as travel time decreases. This approach can further be discounted due to the impact on ΔV requirements during the Neptune capture, given the much higher fly-by velocity at the planet.

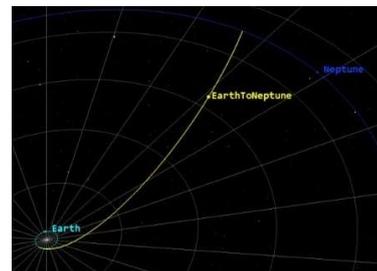

Fig. 5. Direct transfer from Earth to Neptune

Regardless of whether or not there is a launcher capable of reaching this velocity, the goal is to minimize travel time, cost, and mass. To accomplish this, there are various alternative trajectories that have been proven in previous studies to reduce mass and travel time [39]. For example, the simplest trajectory is to use Jupiter's large gravitational field in an Earth-Jupiter-Neptune (EJN) manoeuvre to passively add velocity to the spacecraft through a 'gravitational assist'. Due to the nature of the mission, it may be more appropriate to use additional planetary fly-bys to further reduce fuel needs. An alternative is to reduce the spacecraft's heliocentric velocity and fly by Venus, which would push us back to Earth, and therefore acquire more velocity at this point than would have been provided by a launch vehicle. This Earth-Venus-Earth-Jupiter-Neptune (EVEJN) trajectory can be viewed as a similar approach to the Cassini EVEEJS route to Saturn. Remembering the relatively long orbital periods of Neptune (165 years) and Jupiter (12 years), we can treat Neptune's position as relatively fixed for first-order calculations, meaning our launch window is constrained only by Jupiter and Neptune's location with respect to one another.

All trajectory options have pros and cons and as with any system, complexity gives rise to problems. Complex trajectories carry a higher risk of missing our target, voiding all primary objectives and resulting in a failure of the mission. Additionally, trajectories that take the mission closer to the inner Solar System bring the spacecraft into a higher radiation environment, making hardware damage and problematic failures onboard the spacecraft more likely. Furthermore, the increased spacecraft temperatures would again mean more strenuous design constraints, and additional mass only used to support a transfer, potentially cancelling out any gains resulting from the, in theory, more fuel-efficient trajectory.

Table 1. Pros and cons of Earth-Neptune trajectories

|        | Pros                              | Cons                                                                                                               |
|--------|-----------------------------------|--------------------------------------------------------------------------------------------------------------------|
| Direct | The simple, yearly launch window  | - Fuel expensive<br>- Needs secondary propulsion                                                                   |
| EJN    | Decrease in C3                    | - Increase in complexity<br>- Smaller launch window<br>- Jupiter's environment                                     |
| EVEJN  | Minimal C3                        | - Very complex<br>- Very short launch window<br>- Inner Solar System has an increased probability of radiation damage |

However, gravity assists are well understood and have been performed since the very first interplanetary missions. This lowers the risk of error to one which can be easily mitigated, increasingly so with new control algorithms and technology. The radiation issue could be solved using structural elements and redundancy, but this will only further increase mass. A balance between fuel, travel time, cost, and weight constraints must be achieved before coming to a conclusion on what trajectory is best to use

It is necessary to protect the structures and instruments onboard Somerville and Bingham. Whipple bumper shields can be used to counteract the majority of impacts however covering the entirety of the spacecraft in this type of impact protection. Surface-mounted instruments which would fail when impacted by particles with sizes of 1-10$\mu_m$ must be protected. This primarily covers instruments with exposed sensor arrays or large apertures. Where possible, these instruments can be housed inside of the Somerville-Bingham structures. The remaining exposed instruments can make use of stuffed Whipple shields. This is preferable compared to multi bumper-layer Whipple shields as for the same level of protection, the stuffed Whipple shield will have a reduced profile. Therefore, packaging the spacecraft within the chosen launch vehicle will not require payload volume upgrades.

A major site that could be damaged by IDP impacts would be the high gain dish antenna located on Somerville. As a combined communications and radio science tool, it is relatively important to the normal and emergency operation of the spacecraft. However, protecting this antenna using Whipple bumper shields would lead to large mass and volume penalties. To account for this, thickening the material used for the dish element of the antenna is one of the simplest means to mitigate IDP impacts. In the event that the dish is punctured, data or communication packets received from the spacecraft will be offset by a factor. However, this can be accounted for provided that there are samples of similar data or communications packets received during normal operation to compare to.

*4.2 Propulsion*

Electric propulsion for transfer to Neptune has been discounted, the main reason being the high power demand. As the majority, and by far the most important, region of spacecraft operations will be the outer Solar System, solar power is not feasible as a primary power source. The only viable supply is any kind of nuclear process, or more specifically nuclear fission or nuclear decay. Nuclear reactors were discounted due to both safety and complexity concerns. Radioisotope Thermal Generators (RTGs) however are a tested, reliable and safe means of power generation, but their output is limited. In order to make electric propulsion a real alternative to chemical propulsion (in terms of transfer time), the power available needs to be high and therefore a large number of RTGs be carried. Electric propulsion was also discounted due to reliability concerns. In the optimal case, the propulsion and power system has to work for the entire transfer, around 10 to 15 years. Prolonged operations like these are still untested and therefore risky.

For these reasons, chemical propulsion was selected. More specifically storable, hypergolic propellant, although the exact fuel is still open for consideration. We decided against other propellant combinations as these use at least one cryogenic component. Decade long storage of cryogenic fuel in space is still in its infancy and poses a development risk. Another mark against cryogenic propellant lies in the issue of fueling. Just as it is difficult to store cryogenic fuel for a long time in space, it is difficult and dangerous to store such propellant on Earth during launch vehicle integration and preparation. SpaceX's documentation, though admittedly limited, does not present the option of fueling on the launch pad, so it should be assumed that any fueling needs to be done before launch vehicle integration. A similar situation occurred with the Shuttle-Centaur Program, causing numerous technical and safety issues, ultimately terminating the project [40].

Solid propulsion was dismissed as it is not controllable and we expect many manoeuvres to require high precision, in addition to the relatively low specific impulse.

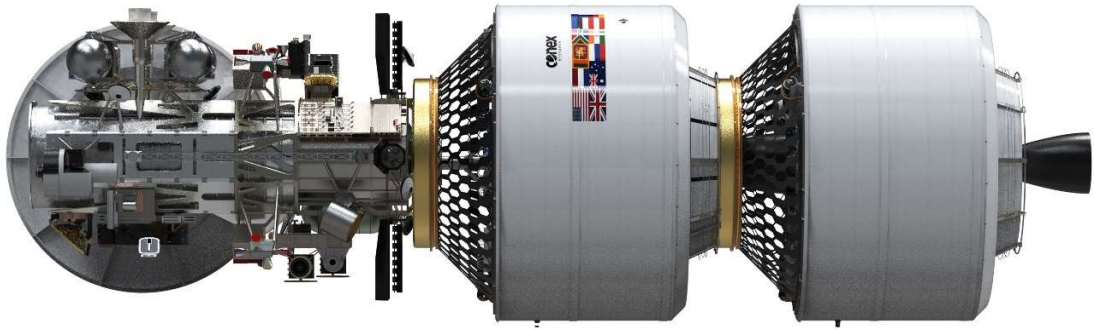

Fig. 6. Arcanum in a CF3 configuration

During the analysis and design of the spacecraft and its transfer to Neptune, a number of feasible configurations for the 'booster stage' – this being the propulsion system which transfers Somerville-Bingham from Earth parking orbit to its destination highly eccentric inclined Neptune orbit, or Poseidocentric orbit – were found. These are presented here. As the project continues to develop, and further constraints are placed on the mission such as financial and political considerations, a single one of these can be downselected for further development of the mission.

The configurations presented below should be viewed in the context of the Neptunian transfer having three distinct phases:
- Earth departure: from an initial geocentric orbit to the interplanetary heliocentric transfer orbit
- Mid-course corrections
- Neptune arrival: from the heliocentric transfer orbit into a highly eccentric inclined Poseidocentric orbit

It is planned that Starship will bring a vehicle and attached propulsion system capable of these three phases into LEO. While it is understood that Starship in a fully expendable configuration may even be capable of interplanetary transfer burns before releasing payloads, this extrapolates too much from the currently available information on Starship capabilities.

*4.2.1 Configuration 1 - restartable booster stage*
Configuration 1, or 'CF1', utilizes one large booster stage with a restartable engine. This stage performs all major manoeuvres.

Once the Neptune Arrival Manoeuvre has been performed, Somerville-Bingham will be released. Both the spacecraft and the recently discarded booster stage will now be in an orbit around Neptune. One option from here could be the reduction of booster perigee for disposal using Neptune, but retention of the booster, under control, in a Neptune orbit for Triton science, discussed later, is also possible.

Advantages: Only one stage requires development and operation as opposed to two, reducing complexity. The reuse of one single propulsion system for all high ΔV manoeuvres is also mass efficient. This also reduces the propellant and propulsion system mass which is integrated into the Somerville spacecraft, making the vehicle more efficient during operation around Neptune.

Disadvantages: The booster stage will be fairly large and therefore difficult to design, build and test. For many years the propellant tank is only partly filled because a large proportion will be used for Earth departure. Any movements and fluid behaviours should be considered and a reliable restart ability after around 15 years needs to be ensured. An eventual discard of the stage will need to be made to avoid space debris or contamination.

*4.2.2 Configuration 2 – non-restartable Earth departure stage*
CF2 also utilizes one booster stage, but in contrast to CF1, this rocket stage is used for the Earth departure burn only. Course corrections and the Neptune capture burn would be done by Somerville's integrated propulsion. No rocket stage reaches Neptune and Somerville arrives alone.

Advantages: The booster stage is only used once within a short period of time from launch. The engines do not need a restart capability and can be designed to be simpler and more reliable. The booster is also smaller and therefore easier to design and test. Fluid behaviour is easier to control because the tanks are being completely emptied. Little changes would be required for Somerville, given it already has to perform complex manoeuvres and in any case needs a restartable engine. In summary, using it for mid-course corrections and Neptune arrival does not require any functionality Somerville's engine don't already possess.

Disadvantages: Somerville becomes much larger as it has to carry the fuel for Neptune arrival integrated into its propulsion system. The danger of leaving space debris in the form of the booster stage is not avoided, only removed from the Neptunian system and made perhaps more unpredictable in a heliocentric orbit. Collision with Solar System bodies of interest would need to be avoided.

*4.2.3 Configuration 3 – non-restartable Earth departure stage and restartable booster stage*

CF3 combines the previous CF1 and CF2: One stage for Earth departure and one stage for mid-course corrections and Neptunian arrival. As in CF1 both Somerville and the booster stage end up in a Neptune orbit.

Advantages: Somerville does not need large amounts of fuel reducing its size, and the problem of a large rocket stage is split into two smaller, more manageable ones. The first stage can be kept simple as it only burns once. For both stages, fluid movements are a lesser issue, because the first stage is emptied in one burn and for the second stage the course-corrections only require a relatively little amount of fuel, leaving the fuel tanks mostly filled for the Neptune capture burn.

Disadvantages: Two smaller stages are heavier than one larger one. Two sets of engines are necessary, in contrast to only one. The fuel tanks are relatively light and scale appropriately. Two stages mean one additional space element, one more to build, integrate and test. One more interface. Complexity is increased.

*4.3 Mission Timeline*

The Arcanum Mission can be divided into six phases as detailed in Fig. 7.

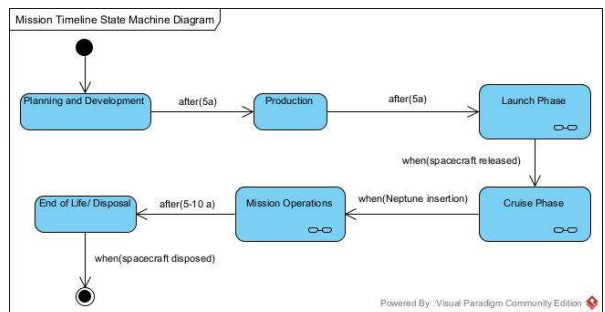

Fig. 7. Arcanum Mission Timeline

Initially, the planning and development of the mission are required, the current phase. Following this a production phase, where building and testing of the spacecraft take place. 10 years can be assumed for each of these, taken from the approximate lengths of similar work on previous missions, discounting outliers but recognising that overrunning is always a possibility.

The operation of the mission begins with the launch phase, shown in Fig. 8. This consists of the integration of the spacecraft with the launcher, pre-launch preparations and the launch and the release of the spacecraft into the targeted orbit; all the responsibility of a launch provider.

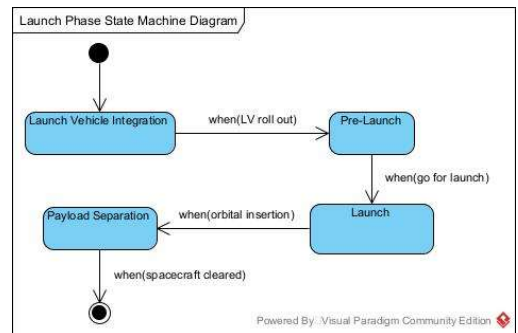

Fig. 8. Launch Phase Timeline

Once the spacecraft has been released, the cruise phase begins. As shown in Fig. 9, this includes Earth Departure, the Jupiter gravity assist and Neptune Arrival. During the years-long interval between these steps, observations of the interplanetary space are performed, detailed previously. Between Earth and Jupiter, the asteroid belt is of interest and beyond Jupiter, centaur objects. No specific targets have been identified as of yet, as this is dependent on the chosen launch window. The telescope payload will increase the scientific range considerably.

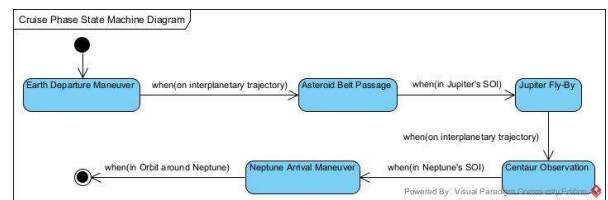

Fig. 9. Cruise Phase Timeline

The exact duration of the cruise phase is still undetermined but is constrained to between 10 and 15 years. After this duration, the actual operational phase begins with arrival at the Neptune system, as detailed in Fig. 10. At this preliminary stage of mission development, two sub-phases have been identified: Phase 1 begins with a fly-by and survey of Nereid, a moon that is on a highly eccentric and high orbit around Neptune.

The main goal of the first phase is to complete a survey of Triton. 40% of its surface is already known [41] and preliminary landing zones are selected, but if these landing zones prove to be inadequate or inferior to others located on the still unknown hemisphere, better landing zones may be chosen. The deployment of the surface probes occurs on this sub-phase and end Phase 1.

For Phase 2 Somerville enters a highly eccentric orbit around Neptune and switched between two science foci: while on the orbit section near Neptune, the focus lies in Neptune, its moons and rings. At the apoposeideum, the science focus lies in telescopic observations of KBOs and other far-away objects.

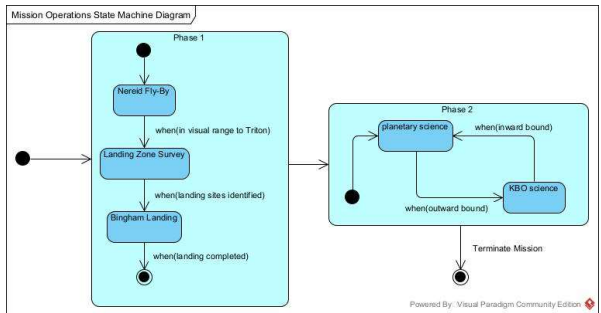

Fig. 10. Mission Operations Timeline

The operational phase is planned for 5 years, but mission extension to 10 or more years is expected. Even the most robust components eventually fail and gradually render whole sub-systems or even Somerville inoperable. To avoid a collision with a moon and potential contamination, the Arcanum mission will be ended with the disposal phase, where Somerville is either placed on a high, stable orbit around Neptune or deorbited in a controlled fashion.

*4.4 Communications*

The primary communications link to and from Earth will be via optical communications, also known as laser. Optical communication is a relatively new technology, but prototypes already exist and experiments have been conducted [42]. Advantages over radio-based communication are increased data rates (up to 1 Mbit/s) at a lower operating power.

A backup method of traditional radio-based communications will be installed on the spacecraft and provide data transfer rates of at least 2 kbit/s. This is sufficient to continue mission operation at a reduced capacity in the eventuality laser communications fails, demonstrated previously by Galileo's reliance on a 100 bit/s low-gain connection following a high-gain failure.

Communication between Somerville and the sub-probes will also be via radio. Somerville's radio communications array will also serve scientific purposes. These include the use of radar and radio occultation experiments. These communications solutions are hosted in a boom-mounted dish similar to a previously proposed solution for Neptune [2].

European Data Relay Satellites (EDRS), equipped with laser terminals for connection with Somerville, are to be used to provide the final link to ground stations using radio. Alternately, ground stations equipped with optical communication equipment can be used to test the reliability of such a direct link across deep space. Back-up communications will use the highly sensitive Deep Space Network (DSN). Fig. 11 shows the architecture of the main communications scenario.

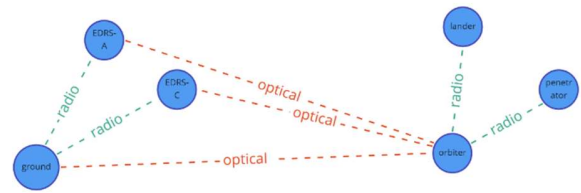

Fig. 11. Primary Communications Architecture

*4.5 Power*

Americium-241 radioisotope thermoelectric generator (RTG) development has been a priority of ESA for several years, and work at the UK National Nuclear Laboratory would be used for the implementation of this new technology in its first deep-space mission [43].

RTGs are the most sensible power generation system to power Somerville given the bulk of the Arcanum Mission takes place around Neptune, where solar power is ineffective. Arcanum is expected to be operational for a minimum of 15 years following transit. The heritage of RTGs used on the previous spacecraft typically uses plutonium-238 as a power source. Furthermore, the power density and half-life of Pu-238 fall within the operational requirements of the Arcanum Mission. However, the global abundance of Pu-238 is very low as it has to be synthesised resulting in Arcanum incurring large costs.

In recent years there has been a shift within Europe where research groups are investigating the use of americium-241 as a power source. In particular the Am-fuelled RTG developed by the University of Leicester as part of the ESA radioisotope power system programme. The proof of concept RTG stack consisted of three Am-241 pellets capable of producing a specific electric power of 1.1 W/kg [44]. As the Technology Readiness Level (TRL) of the RTG is set to increase over the next decade, it is not unreasonable to assume a more energy-dense Am-241 RTG stack will be available by the date of Arcanum launch. Therefore, it is assumed that each stack will have a specific electrical power of 1.5 W/kg.

The use of Am-241 will also mitigate concerns laid out in the most recent NASA decadal survey [3], which expresses concerns that a lack of domestic Pu-238 production in the United States and reliance on Russia presents problems for future such missions.

A number of options are available for the storage of power generated by the RTGs. Hydrogen fuel cells, for example, would provide a high capacity, but in turn, require a high level of thermal control. Lithium-ion cells offer a high promise in space exploration due to high reliability, high energy density, a high number of charge/discharge cycles, no start-up delay and little to no passive discharge, making them ideal for long-duration journeys. However, they are less tested in this context, exhibit some instability at high temperature and do suffer damage when held at low charge levels. Other alternatives such as lithium-sulfur dioxide, lithium sulfide, nickel-cadmium, nickel-hydrogen and silver-zinc were considered, each with respective advantages and disadvantages [45].

In conclusion, however, the high reliability and rechargeability of lithium-ion cells when in extreme environments, when compared to legacy battery solutions, made this the power storage solution of choice for Arcanum. Due to Somerville's onboard RTG system, the spacecraft will be designed to sustain itself with a low power capacity. This means that the only time the spacecraft would require electric charge from storage is for higher voltage applications like scientific data transfer/relay or instrument operation; lithium-ion batteries excel in this use case as they do not get damaged or passively discharge when at high charge capacities, therefore holding their charge reliably for years.

Lithium-ion batteries also have high energy density and can, therefore, lower the amount of overall spacecraft mass.

Lithium-ion cells are able to provide almost instantaneous power from startup, with no extra "warm-up" time to get to a sufficient operational voltage, making this a great pairing with RTG's to be able to better sustain constant system operation, even in high load scenarios like science transfer, with minimal extra power management systems when compared with silver zinc.

*4.6 Bingham*

As aforementioned, due to the scope of this study and the detail into which Somerville is described, further information on the Bingham soft-landing probe and surface penetrators will be made available in subsequent conference papers. However, a top-level summary is briefly mentioned here.

The duration of the Triton-surface segment of this mission is dictated by science requirements. Trade-offs between science requirements and engineering constraints have arrived at an approximate 6-month duration, with this only constrained by component lifetime, given the inclusion of RTG power and an, in effect, lifetime unconstrained by power.

Of all the worlds where a soft landing has been attempted, Earth's Moon is most similar to Triton. Differences relevant for a long-term operation like composition and the ambient temperature are present, but for landing, two other factors are important: the lack of an atmosphere and the gravity. Mars is dissimilar to Triton as it has an atmosphere that is thin, but not thin enough to be entirely neglected. Furthermore is sustains a much higher surface gravity. Asteroids on the other hand, while devoid of an atmosphere as well, have almost no gravity at all. For this reason, we look for the landing methods of lunar probes.

The most common choice is propulsive landing via rockets. Most landers have used storable bipropellant engines, with the exception of Starship and Blue Moon. These two still conceptual landers use cryogenic propellant.

The second alternative was used in the very first lunar landers but has since then not been utilized again: soft landing via airbags to compliment a propulsive landing. This has been selected for Bingham given a requirement to avoid as much contamination of the landing site as possible. These airbags would then deflate, leaving a rigid connection between the spacecraft structure and Triton's surface, avoiding any damping of any seismometer measurements.

*4.7 Penetrators*

As with Bingham, details of penetrator operation are reserved for future publications. However, these can be seen as analogous to the Mars96 Mission [46], augmented with more modern technology.

## 5. Discussion

The considerations of such a new kind of proposal for the outer Solar System are broad and wide-ranging, thus limiting the scope of this single conference paper. Nonetheless, it is clear that the necessity of such a mission can be demonstrated, and the paradigm shift attempted with such a proposal is of interest to the discussion around a new generation of launchers. The definitive publication of the Phase A study of this mission will be forthcoming and in the meantime selected topics, such as those covered in this paper, attempt to both raise awareness of the feasibility of such missions with SHLLVs and demonstrate one possible concept for a comprehensive science platform at Neptune.

**Acknowledgements**

This work has been completed by the research group Conex (Appendix A). This virtual-working group has received tremendous and dedicated efforts from every member of the team, and consistent support from the founding members. Thanks are given to all team members and every additional external supporter, speaker and technical reviewer. As Conex welcomes Sydereal Pte Ltd as a new partner, our thanks are also extended to them for their support of our work.

The international nature of Conex has proved particularly challenging as COVID has impacted different locations at various times. The strength and unity of our members however have been steadfast, and nothing short of admirable.

**Appendix A (Conceptual Exploration Research, 'Conex')**

Conex Research, a portmanteau of Conceptual and Exploration, was founded in April 2020 during the first COVID-19 lockdown as a platform for early-career professionals - current students, graduates and under 30s

- to develop their skills in space research and proposal writing [47]. This group has now expanded to encompass members from six continents, with experience in numerous science and engineering disciplines, as well as the graphics and management domains.

Conex is growing rapidly in membership, a direct result of establishing a strong presence on social media and openly sharing progress. The group also provides training and mentoring, through its partners, in Space Mission Design, Operations and Project Management.

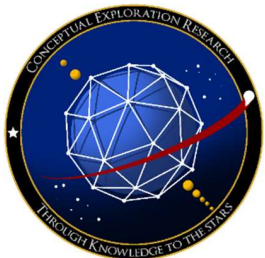  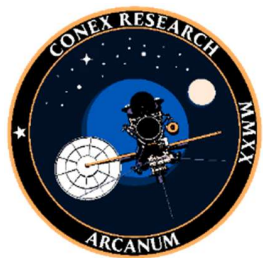

Fig. A1. Conex Seal    Fig. A2. Arcanum Mission Patch